\documentclass[aps, pre, twocolumn, longbibliography, amsfonts, amsmath, amssymb, 10pt,x11names]{revtex4-1}
\usepackage[T1]{fontenc}
\usepackage{graphicx}
\usepackage{bm}
\usepackage{hyperref}
\hypersetup{colorlinks=true, linkcolor=red, citecolor=blue, urlcolor=blue}
\usepackage{float}
\usepackage{multirow}
\usepackage{times}
\usepackage{tikz}
\usepackage{xcolor}

\begin{document}

\definecolor{ma}{rgb}{1, 0, 1}
\definecolor{vi}{rgb}{0.58, 0, 0.827}
\definecolor{gr}{rgb}{0.196, 0.804, 0.196}
\definecolor{br}{rgb}{0.647, 0.165, 0.165}
\definecolor{cy}{rgb}{0, 1, 1}

\title[]{Phase separation of a magnetic fluid: Asymptotic states and non-equilibrium kinetics}  
\author{Anuj Kumar Singh}
\author{Varsha Banerjee}
\affiliation{Department of Physics, Indian Institute of Technology Delhi, New Delhi, 110016 India}

\begin{abstract}   
We study self-assembly in a colloidal suspension of magnetic particles by performing comprehensive molecular dynamics simulations of the Stockmayer (SM) model which comprises spherical particles decorated by a magnetic moment. The SM potential incorporates dipole-dipole interactions along with the usual Lennard-Jones interaction and exhibits a gas-liquid phase coexistence observed experimentally in magnetic fluids. When this system is quenched from the high-temperature homogeneous phase to the coexistence region, the non-equilibrium evolution to the condensed phase proceeds with the development of spatial as well as magnetic order. We observe density-dependent coarsening mechanisms - a diffusive growth law $\ell(t)\sim t^{1/3}$ in the nucleation regime, and hydrodynamics-driven inertial growth law $\ell(t)\sim t^{2/3}$ in the spinodal regimes. [$\ell(t)$ is the average size of the condensate at time $t$ after the quench.] While the spatial growth is governed by the expected conserved order parameter dynamics, the growth of magnetic order in the spinodal regime exhibits unexpected non-conserved dynamics. The asymptotic morphologies have density-dependent shapes which typically include the isotropic sphere and spherical bubble morphologies in the nucleation region, and the anisotropic cylinder, planar slab, cylindrical bubble morphologies in the spinodal region. The structures are robust and nonvolatile, and exhibit characteristic magnetic properties. For example, the oppositely magnetized hemispheres in the spherical morphology impart the characteristics of a {\it Janus particle} to it. The observed structures have versatile applications in catalysis, drug delivery systems, memory devices, and magnetic photonic crystals, to name a few.

\end{abstract}
\maketitle

%%%%%%%%%%%%%%%%%%%%%%%%%%%%%%
\section{Introduction}

Magnetic fluids, also referred to as {\it ferrofluids}, are comprised of single-domain magnetic particles dispersed in a carrier liquid \cite{Odenbach2002, Rosensweig1985, Huke_2004}. The magnetic inclusions interact via dipole-dipole interactions which are anisotropic and long-range. They form aggregates of unusual shapes that exhibit magnetic order even in the absence of external fields \cite{Meyer1975, Wei1992, Weis1992, Groh1994, Gao2000, Klapp2002, Stevens1995}. The application of an external field provides additional control that can be used to switch or modify the statistical states of the fluid \cite{Stevens1995, Samin2013, Salzemann2009}. The twin properties of fluidity and magnetism impart unique characteristics, with many open questions in their equilibrium and non-equilibrium behavior. So there is need to address them from the point of view of fundamental physics and promising applications. For example, self-assembled magnetic chains and rings are being used as a basis to understand equilibrium polymerization, dynamic heterogeneities in glass formers, loopless branched structures and gels, phase behavior in network fluids, etc. \cite{Semlyen2005, Biroli2022, Sciortino2009, Dias2017}. Compact aggregates on the other hand are also capturing great interest because of their high surface to volume ratio, large pore volume, and low density \cite{Ulusoy2023}. These features are being exploited in controlled encapsulation-release of drugs and medical diagnostics, energy storage, and conversion, creation of molecular biomaterials such as fibers and tubes, the building of nano-structures and nano-devices, etc. \cite{Jia2020, Li2022, Zahn2001}.   

Nonpolar fluids have been widely modelled by the Lennard-Jones (LJ) potential that includes a repulsive and attractive term \cite{Allen2017, Nicolas1979}. Particles of a dipolar fluid have a magnetic moment, so it is imperative to include dipole-dipole interaction in addition to the LJ potential for their theoretical studies. The Stockmayer (SM) model has been popularly used to capture the essential features of magnetic fluids, namely the observation of the gas-liquid (GL) co-existence phase in the density-temperature ($\rho-T$) space \cite{Smit1989, Van1993, Leeuwen_1993, Dongsheng_1995, Bartke_2007, Kalyuzhnyi2007, Richardi_2008, Richardi_2009, Moore2015}. Understanding the GL diagram of fluids has been an important topic of study. The coexistence regime of the LJ fluid is well-studied via Monte Carlo (MC) and molecular dynamics (MD) simulations \cite{Smit1992, Watanabe2012, Heyes2015, MacDowell_2006, Schrader2009, Block_2010, Binder_2012}. They have revealed that the competition between the various terms in energy yields condensates with distinct shapes, e.g., spheres, cylinders, planar slabs, etc., for different values of $\rho$.  Similar but sporadic observations have also been made in the context of the SM fluid. In addition to having distinct shapes, we can also anticipate in them a characteristic magnetic order due to the anisotropic dipole-dipole interactions. Although the dual characteristics of self-assembly and magnetism have exciting fundamental physics and technological applications, a methodical study identifying the density intervals for the different self-assembled shapes, the influence of temperature on the internal structure, and their magnetic properties are missing in the literature. In the present paper, we fill these lacunae with the help of comprehensive MD simulations of the SM model.  
 
Our starting point is a quench from the high-temperature isotropic gas phase into the GL coexistence region. The simulations have been performed for a SM gas with a prototypical value of magnetic moment $\mu=2.5$ with a critical point $\rho_c = 0.29(1)$, $T_c = 2.63(1)$ in LJ units. In a recent paper \cite{Singh2023}, we studied the approach to equilibrium in the {\it spinodal} regime and observed an inertial growth $\ell_s(t) \sim t^{2/3}$, where $\ell_s(t)$ is the typical lengthscale of the condensate at time $t$. Though predicted by Furukawa in 1985 \cite{Furukawa1985}, the inertial growth law was never observed in MD simulations, and this made our observation significant. Additionally, we find that the magnetic order is triggered after the onset of condensation (or spatial order). The present paper focuses on the asymptotic structures ($t\rightarrow \infty$) as a sequel to our non-equilibrium studies. We concentrate on the structural and magnetic properties of these morphologies by a systematic variation of parameters in the $\rho-T$ plane. The main observations from our paper are as follows: 
(i) The non-equilibrium evolution is distinct in the nucleation and spinodal regimes. The asymptotic structures have density-dependent shapes with characteristic spatial and magnetic order. They are isotropic in the nucleation regime but anisotropic in the spinodal regime. 
(ii) We identify density intervals which yield a sphere, cylinder, slab, cylindrical bubble, and spherical bubble. Naturally, these are minimum energy shapes for the corresponding densities.  
(iii) The magnetic moments always co-align with the surface. Consequently, the morphologies have unusual magnetic features even in the absence of external fields. For instance, the sphere is a {\it magnetic Janus particle} due to oppositely magnetized hemispheres. The spherical bubble on the other hand has large magnetization on the surface which gradually reduces at the center. 
(iv) The anisotropic condensates are comprised of dipole chains along the direction of anisotropy. They exhibit perfect magnetic order ($ M \simeq 1$) even in the liquid state ($T \simeq 1.05$) as the dipole-dipole interactions overwhelm the thermal energy. (v) For lower temperatures ($T\simeq 1.0$), the structures solidify, and exhibit quasi-long-range order which is predominantly  face-centered. (vi) At still lower temperatures  ($T\lesssim 0.9$), there is neither long-range spatial nor magnetic order. The aggregates are glassy with $ M \simeq 0$ and an Edwards-Anderson spin-glass order parameter $q_{EA}\simeq 1$. 

In what follows, we will focus on understanding the interplay of the short-range steric repulsion and long-range dipolar interactions along with the inherent magnetism of the SM particles in the self-assembled condensates. Our paper is organized as follows. Section~\ref{sec:2} provides the model and tools for characterizing the structural and magnetic properties of the SM condensates. The simulation details and numerical results are provided in Sec.~\ref{sec:3}. These include a comprehensive analysis of the structural and magnetic properties of the condensates for a range of temperatures to access the liquid, solid, and glass phases. In Sec.~\ref{sec:4}, we provide observations of the non-equilibrium evolution of spatial and magnetic order in the nucleation regime. Finally, Sec.~\ref{sec:5} contains the conclusion with a summary of results and discussion.

\section{Model and Methodology}
\label{sec:2}
\subsection{Stockmayer model}
The SM model mimics fluids composed of spherical particles with magnetic moments embedded at their center. Let us consider $N$ particles with magnetic moment ${\vec{\mu}} = \mu\hat{\mu}$. The SM potential between particles $i$ and $j$ separated by a distance $\vec{r_{ij}} = r_{ij}\hat{r_{ij}}$ is given by \cite{Leeuwen_1993}: 
\begin{eqnarray}
\label{SM}
U(\vec{r_{ij}}, \hat{\mu_i}, \hat{\mu_j}) = 4\epsilon\sum_{i,j}\bigg[{\bigg( \frac{\sigma}{r_{ij}}\bigg)}^{12}-{\bigg( \frac{\sigma}{r_{ij}}\bigg)}^6\bigg] \nonumber\\
+\frac{\mu_0 \mu^2}{4\pi}\sum_{i,j}\bigg[ \frac{\hat{\mu}_i.\hat{\mu}_j - 3(\hat{\mu}_i.\hat{r}_{ij})(\hat{\mu}_j.\hat{r}_{ij})}{r_{ij}^3} \bigg].
\end{eqnarray}
The first two terms correspond to the LJ potential that describes the short-ranged steric repulsion and the weak van der Waal's attraction. The parameters $\sigma$ and $\epsilon$  are the particle diameter and depth of the attractive potential. They set the spatial and energy scales in the system. The third term represents the dipole-dipole interactions which are significant up to large distances and can be 0 or $\pm$, depending on the position and orientation of the dipoles $i$ and $j$. Clearly, the head-to-tail orientation of dipole moments has maximum attraction while head-to-head orientation corresponds to maximum repulsion. The perpendicular orientation of dipole moments is equivalent to the LJ potential. The SM particles thus experience isotropic short-range interactions as well as anisotropic long-range dipolar interactions. 

The condensation of a liquid drop from the vapor phase and its subsequent non-equilibrium growth is a problem of utmost importance in phase transformations. An intriguing aspect of dipolar fluids is the existence of a magnetic fluid phase in the absence of an applied field \cite{Meyer1975, Wei1992, Weis1992, Groh1994, Gao2000, Klapp2002, Stevens1995}. Consequently, the GL co-existence region of magnetic fluids is a topic of much theoretical and experimental research.  When cooled below the critical temperature $T_c$, the SM model exhibits a phase transition from a paramagnetic gas phase to a GL co-existence phase. Other models that have been popularly used to study magnetic fluids comprise of dipolar soft spheres or dipolar hard spheres which include only the steric repulsion part \cite{Rovigatti2012, Hynninen2005}. These models favor the formation of chains, but they do not exhibit GL phase coexistence that has been observed experimentally in magnetic colloids and ferrofluids \cite{Dubois2000, Mamiya2000, Ivanov2020}. The SM model on the other hand captures the essential features of magnetic fluids and is therefore more representative of them. The GL phase diagram in the $\rho-T$ plane is conventionally obtained from the equation of state \cite{Schmelzer2019}, but computer simulations provide an alternative route, especially when interactions are complex. The coexistence phase diagram of the SM fluid is well-studied and is believed to occur for all dipole strengths. It has been determined for a range of $\mu$ values using MC and MD simulations \cite{Smit1989, Van1993, Leeuwen_1993, Bartke_2007, Kalyuzhnyi2007, Dudowicz2004, Leeuwen1994, Hentschke2007}. The primary effect of increasing $\mu$ is to shift the critical point $(\rho_c, T_c)$ upwards, thereby enlarging the GL co-existence region. 

\subsection{Methodology}
When a system is quenched from high-temperature disordered state ($T>T_c$) to a low-temperature ordered phase ($T<T_c$), there is a formation of the condensed phase that coarsens with time to yield another asymptotic state. As the SM fluid has both spatial and magnetic order, we use a variety of tools to understand the non-equilibrium evolution and the organization in these self-assemblies. 

\subsubsection{\textbf{Pair correlation function}}
\label{sec:PCF}
The standard probe to envisage the internal arrangements of particles within the condensed phase is the pair correlation function (PCF). It measures the probability of finding two molecules separated by distance $r$ relative to that in an ideal gas: $g(r) = \langle \overline{\rho(r)}\rangle/\rho_0$ where $\rho_0=N/V$ is the density of the ideal gas and $\overline{\rho(r)}$ is the average density of the system around $r$. The numerical evaluation is facilitated by the following formula \cite{Weis1993, Birdi2022}:
\begin{equation}
\label{PCF}
g(r)=\frac{1}{N\rho_0}\bigg\langle\sum_{\stackrel{i,j}{i\neq j}}^N\frac{\delta(r-r_{ij})}{(4/3)\pi[(r+\Delta r)^3-r^3]}\bigg\rangle.
\end{equation}
The $\delta$- function is unity if $r_{ij}$ falls within the shell centered on $r$ and is zero otherwise. The division by $N$ ensures that $g(r)$ is normalized to a per-particle function. By construction, $g(r)=1$ for an ideal gas, and any deviation implies correlations between the particles due to the inter-particle interactions. In the liquid phase, $g(r)$ exhibits a large peak at small-$r$ signifying nearest neighbor correlations followed by small oscillations which eventually approach 1 at large-$r$. (The latter signifies loss of correlations at large-$r$). The solid phase is characterized by several sharp peaks at values of $r$ that correspond to the lattice spacing of crystal structures. 

A natural evaluation in the context of the SM fluid is the magnetization which measures the alignment of dipoles: ${\bf M} = \sum_{i=1}^{N} \vec{\mu_i}/N = M\hat{\bf m}$. A perfect ferromagnetic order corresponds to $M=1$, and the paramagnetic or disordered state is characterized by $M=0$. In the presence of an anisotropy direction such as $\hat{m}$, it is appropriate to evaluate the directional PCF \cite{Weis1993}:
\begin{align}
g_{\parallel}(r_{\parallel})=\frac{1}{N\rho_0}\bigg{\langle}\sum_{\stackrel{i,j}{i\neq j}}^N\frac{\delta(r_{\parallel} - r_{ij, \parallel})\theta(\sigma/2-r_{ij,  \bot})}{\pi (\sigma/2)^2 h}\bigg{\rangle}, 
\label{gpar}
\end{align}
where $ r_{ij, \parallel}= |\vec{r_{ij}}.\hat{m}| $ is the separation of particles along the direction of the anisotropy (magnetization) axis and $ r_{ij,  \bot} = |\vec{r_{ij}}- (\vec{r_{ij}}.\hat{m})\hat{m}| $ is the separation in the perpendicular direction. The step function $\theta(x)$ ensures that the cylinder of radius $r=\sigma /2$ has a height $h$ that is used for the discretization of the simulation box. The PCF in the perpendicular direction, $g_{\bot}(r_{\bot})$, can be evaluated analogously.

\subsubsection{\textbf{Bond order parameter}}
\label{sec:BOP}
The local crystalline order in undercooled liquids and solids can be conveniently obtained using the local bond order parameter (BOP) $q_4$ and $q_6$ evaluated from Refs. \cite{Steinhardt1983, Errington2003, Shrivastav2021, Gasser2014}:
\begin{align}
\label{q1}
q_l(i)=\sqrt{\frac{4\pi}{2l+1}\sum_{m=-l}^{l}|\Bar{q}_{lm}(i)|^2},
\end{align}
with
\begin{align}
\label{q2}
\Bar{q}_{lm}(i)=\frac{1}{N_n(i)+1}\sum_{k=0}^{N_n(i)}q_{lm}(k),
\end{align}
and 
\begin{align}
\label{q3}
q_{lm}(i)=\frac{1}{N_b(i)}\sum_{j=1}^{N_b(i)}Y_{lm}(r_{ij}).
\end{align}
In Eq.~(\ref{q2}), $N_n(i)$ includes all the nearest neighbors of particle $i$ [$=N_b(i)$ in Eq.~(\ref{q3})], and the particle $i$ itself. $ Y_{lm}(r_{ij})$'s are the spherical harmonics, with $l$ as a free integer parameter and $m =-l,\cdot\cdot,l$. The BOPs or the $ q_l(i)$'s have characteristic values for different structures and are indicated in Table~(\ref{q}). 

\begin{table}[H]
\caption{Values of $q_4$ and $q_6$ for the perfectly symmetric configurations \cite{Steinhardt1983, Gasser2014}.}
\begin{ruledtabular}
\begin{tabular}{ccc}
Structures & $q_4$ & $q_6$ \\
\hline
\noindent Simple cubic (SC) & 0.764 & 0.354 \\
Body-centered cubic (BCC) & 0.509 & 0.629 \\
Face-centered cubic (FCC) & 0.190 & 0.575 \\
Hexagonal close-packed (HCP) & 0.097 & 0.484 \\
\end{tabular}
\end{ruledtabular}
\label{q}
\end{table}

\subsubsection{\textbf{Edwards-Anderson order parameter}}
In the coexistence region, the dipolar particles form long chains that co-align to form ordered domains.  At lower temperatures, these domains are smaller and randomly oriented due to freezing of the moments. An appropriate order parameter for capturing the arrangements of dipolar particles inside the local frozen domains is Edward Anderson's (EA) order parameter defined as \cite{Edwards1975, Parisi1983}:
\begin{equation}
q_{\text{EA}}=\left[\langle \mu_i \rangle ^2\right]_{\text{av}},
\label{qEA}
\end{equation}
where $\langle \rangle$ is the thermal or dynamic average that yields a non-zero value for frozen dipolar particles and $[...]_{\text{av}}$ is an ensemble average. In the paramagnetic phase, $q_{\text{EA}}=0$ along with the $M=0$. In the ferromagnetic phase, $q_{\text{EA}}\ne 0$ and $M\ne0$. In the frozen (glassy) phase, on the other hand, $q_{\text{EA}}\ne0$ but $M\simeq0$.

\subsubsection{\textbf{Correlation function}}
\label{sec:cor}
The time evolution of morphologies via domain growth is well-captured by the two-point equal-time correlation function $C(\vec{r}, t)$ defined as \cite{Bray2002, Puri2009}:
\begin{align}
\label{corrfn}
C(\vec{r_i},\vec{r_j}, t)=\langle \psi(\vec{r_i}).\psi(\vec{r_j}) \rangle - \langle\psi(\vec{r_i})\rangle \langle\psi(\vec{r_j})\rangle,
\end{align}
where $\psi(\vec{r})$ is the appropriate order parameter and the angular bracket denotes an ensemble average. If the system is isotropic and is characterized by a unique length scale $\ell(t)$, the correlation function obeys dynamical scaling form \cite{Binder1974}:
\begin{align}
\label{co}
C(r, t) \equiv f(r/\ell),
\end{align}
where $f(x)$ is the scaling function. The characteristic length scale $\ell(t)$ is defined as the distance over which the correlation function decays to (say) 0.5 of its maximum value. Small-angle scattering experiments yield the structure factor, which is the Fourier transform of the correlation function \cite{Puri2009}:
\begin{align}
\label{strucfact}
    S(\vec{k},t) = \int d\vec{r}   e^{-i\vec{k}.\vec{r}} C(\vec{r},t),
\end{align}
where $\vec{k}$ is the wave vector of the scattered beam. The corresponding scaling form is given by:
\begin{align}
\label{st}
S(k, t) \equiv \ell^d g(k\ell).
\end{align}
The tail of the structure factor conveys information about defects in the morphologies. For a $d$-dimensional system with an $n$-component of order parameter, $S(k) \sim k^{-(n+d)}$ as $k \rightarrow \infty $. For $n=1$, the scattering is off smooth interfaces and the corresponding scattering function is called the Porod law \cite{Bray2002, Puri2009}. For $n>1$, the scattering is from the different topological defects such as vortices ($n = 2$, $d = 2$), strings ($n = 2$, $d = 3$), and monopoles or hedgehogs ($n=3$, $d=3$). 

\subsubsection{\textbf{Domain growth laws}}
\label{sec:l(t)}
The growth of the ordered phase or domains proceeds via the annihilation of defects \cite{Bray2002, Puri2009}. The determination of the domain growth law $\ell(t) \sim t$ is an important evaluation in phase ordering experiments as it reveals details of the free-energy landscape and relaxation time scales in the system. For example, phase separating solid mixtures with non-conserved dynamics obey the Lifshitz-Allen-Cahn law\cite{LAC_1962, LAC_1979}: $\ell(t) \sim t^{1/2}$. On the other hand, solid mixtures with conserved kinetics and diffusive transport follow the Lifshitz-Slyozov law\cite{LS_1961}:  $\ell(t) \sim t^{1/3}$. These growth laws are characteristic of systems with no energy barriers to coarsening and a unique relaxation timescale. In phase-separating fluids however, the evolution to the equilibrium state is dominated by capillary forces, viscous dissipation, and fluid inertia. The co-existing phases or domains grow with time via the process of nucleation or spinodal decomposition as the case may be \cite{Puri2009, Debenedetti2000}. In fluids and polymers however, hydrodynamic effects become important after the initial diffusive regime.  A dimensional analysis leads to the following additional growth regimes: $\ell(t)\sim t$ for $\ell(t)\ll \ell_{i}^*$;  $\ell(t)\sim t^{2/3}$ $\ell(t)\gg \ell_{i}^*$. The inertial length scale $\ell_{i}^*=\eta^2/\tilde{\sigma}\rho$, where  $\tilde{\sigma}$ is the interfacial tension, $\rho$ is the fluid density and $\eta$ is the shear viscosity. It marks the cross-over from a low-Reynolds number ($R=\rho/\eta\ell$) viscous hydrodynamic regime to an inertial regime \cite{grant1999}. Domain growth is even more complicated in disordered systems due to the pinning of interfaces at disorder sites and their subsequent roughening. Such systems exhibit logarithmic growth signifying a multitude of lengthscales, energy barriers, and relaxation times  \cite{Emig1998, Puri2004, Manoj2017}.   

\section{Simulation Details and Asymptotic States}
\label{sec:3}
We have performed MD simulations ($d=3$) of the SM fluid in the canonical (NVT) ensemble using periodic boundary conditions. The magnetic particles interact via long-range dipolar interactions, so a truncation of the range of interaction leads to inaccurate results. To prevent this, we repeat the simulation cell periodically and use the {\it Ewald summation technique} \cite{Frenkel2001}. 
So the simulations represent the thermodynamic limit that brings out the effect of the long-range dipole-dipole interactions without interruptions from systemic lengthscales. The long-time evolutions have been performed using the Langevin thermostat which ensures that the temperature of the system has only small fluctuations about the desired fixed value \cite{Davidchack2009}. For the non-equilibrium studies of domain growth, it is essential to incorporate hydrodynamics. We use the Nos\'e-Hoover thermostat for this purpose, which is known to preserve the relevant feature of hydrodynamics for domain growth \cite{nose1984, hoover1985, hoover1999}. The asymptotic structures are statistically identical for either choice of thermostat.  

The MD simulations were performed using LAMMPS  \cite{LAMMPS}. We have taken 4000 particles in a cubic box whose length is adjusted according to the desired system density. The time evolution of positions and velocities of the particles has been implemented using the velocity-Verlet algorithm with simulation time step $\Delta t = 0.002$. All the calculations are performed in reduced LJ units by defining $ T^*= k_BT/\epsilon$, $ \rho^*= N\sigma^3/V $, $ \mu^*= \mu/\sqrt{\epsilon \sigma^3} $,  $ \Delta t^*= \Delta t/\sqrt{m\sigma^3/\epsilon} $, where $N$ is the total number of particles, $V$ is the volume of the simulation box, and $k_B$ is the Boltzmann constant. (The star is dropped in the subsequent discussions.) Starting from a homogeneous gas of SM particles at high temperature $T=5$ (in LJ units), the system is first allowed to stabilize at high temperature and then once again after another quench in the GL coexistence regime. Before making the production runs, we confirm that the system has reached the asymptotic state by ensuring that the energy fluctuations about the mean value are small. Finally, a run of $10^6$ steps was performed, in which the system configurations were measured at intervals of $500$ steps. All the data presented has been averaged over 20 independent samples.

\subsection{The $\rho-T$ plane}
\begin{figure*}
    \includegraphics[width=1\textwidth]{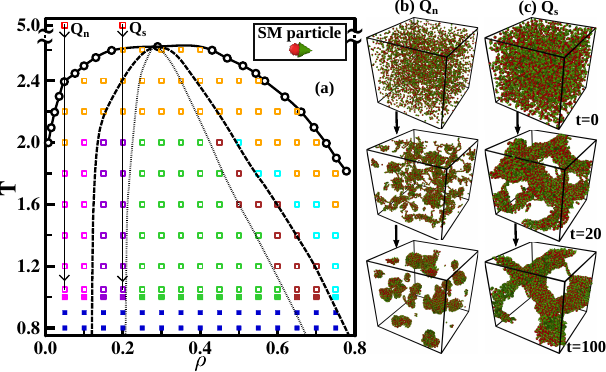}
    \caption{(a) Regions corresponding to different asymptotic structures in the GL coexistence region of SM fluid for $\mu=2.5$. The co-existence curve (solid line) in reduced LJ units is read from Ref. \cite{Stevens1995}. The colored squares indicate the different structures that emerge after a quench from $T=5.0$ to these points: sphere ({\color{ma}\protect \tikz \protect \draw[line width=1.5pt] (0,0) rectangle (0.15,0.15);}), cylinder (\protect \tikz \protect \draw[vi, line width=1.5pt] (0,0) rectangle (0.15,0.15);), planar slab 
    (\protect \tikz \protect \draw[gr, line width=1.5pt] (0,0) rectangle (0.15,0.15);), cylindrical bubble 
    ({\color{br}\protect \tikz \protect \draw[line width=1.5pt] (0,0) rectangle (0.15,0.15);}), spherical bubble 
    (\protect \tikz \protect \draw[cy, line width=1.5pt] (0,0) rectangle (0.15,0.15);), see Fig.~\ref{2} for the asymptotic structures. The orange squares at high temperatures correspond to a nearly isotropic state since the condensed liquid phase here is tiny amounts. The filled blue squares indicate states with random spatial arrangement and magnetic orientation of the dipoles. The dashed and dotted lines demarcating the regions with distinct structures are a guide to the eye.  The arrows from $Q_n$ and $Q_s$ (red squares) represent quenches from $T=5$ to $T=1.05$ in the nucleation region ($\rho =0.05$) and the spinodal region ($\rho =0.2$). (b) Evolution morphologies corresponding to $Q_n$ demonstrating nucleation and subsequent growth by diffusion and coalescence. (c) Evolution morphologies corresponding to $Q_s$ exhibiting bi-continuous structures characteristic of spinodal decomposition.}
    \label{1}
\end{figure*}

Let us refer to the GL coexistence curve shown in Fig.~\ref{1}(a) for $\mu=2.5$. (This data has been read from Ref. \cite{Stevens1995}.) Starting from an initial ($t=0$) isotropic state at $T = 5.0$, let us examine the condensation initiated by quenches to $T=1.05$ for $\rho=0.05$ [${\bf Q_n}$] and $\rho = 0.2$ [${\bf Q_s}$] shown by the arrows. Figure~\ref{1}(b) shows the typical evolution snapshots for ${\bf Q_n}$. Clearly, the growth is via nucleation and subsequent coalescence. We observe a similar scenario for quench points just below the co-existence curve and demarcate this region by the dashed line or the {\it spinodal curve}. Figure~\ref{1}(c) show the evolution for ${\bf Q_s}$. The bi-continuous morphology at intermediate times is typical of phase separation via {\it spinodal decomposition.} This route for phase separation is observed for all data points in the {\it spinodal region} below the spinodal curve. It is important to point out here that in an infinite system, the bi-continuous patterns will persist forever. The coarsening system is affected by the finite size $L$ of the box, and settles into a morphology that does not evolve further in time. The asymptotic patterns that we will discuss arise in this finite-size limit. 

%%%%%%%%%%%%%%%%%%%%%%%%%%%%%%%%%%%%%

%%%%%%%%%%%%%%%%%%%%%%%%%%%%%%%%%%%%%
\begin{figure}
    \begin{center}
    \includegraphics[width=1\linewidth]{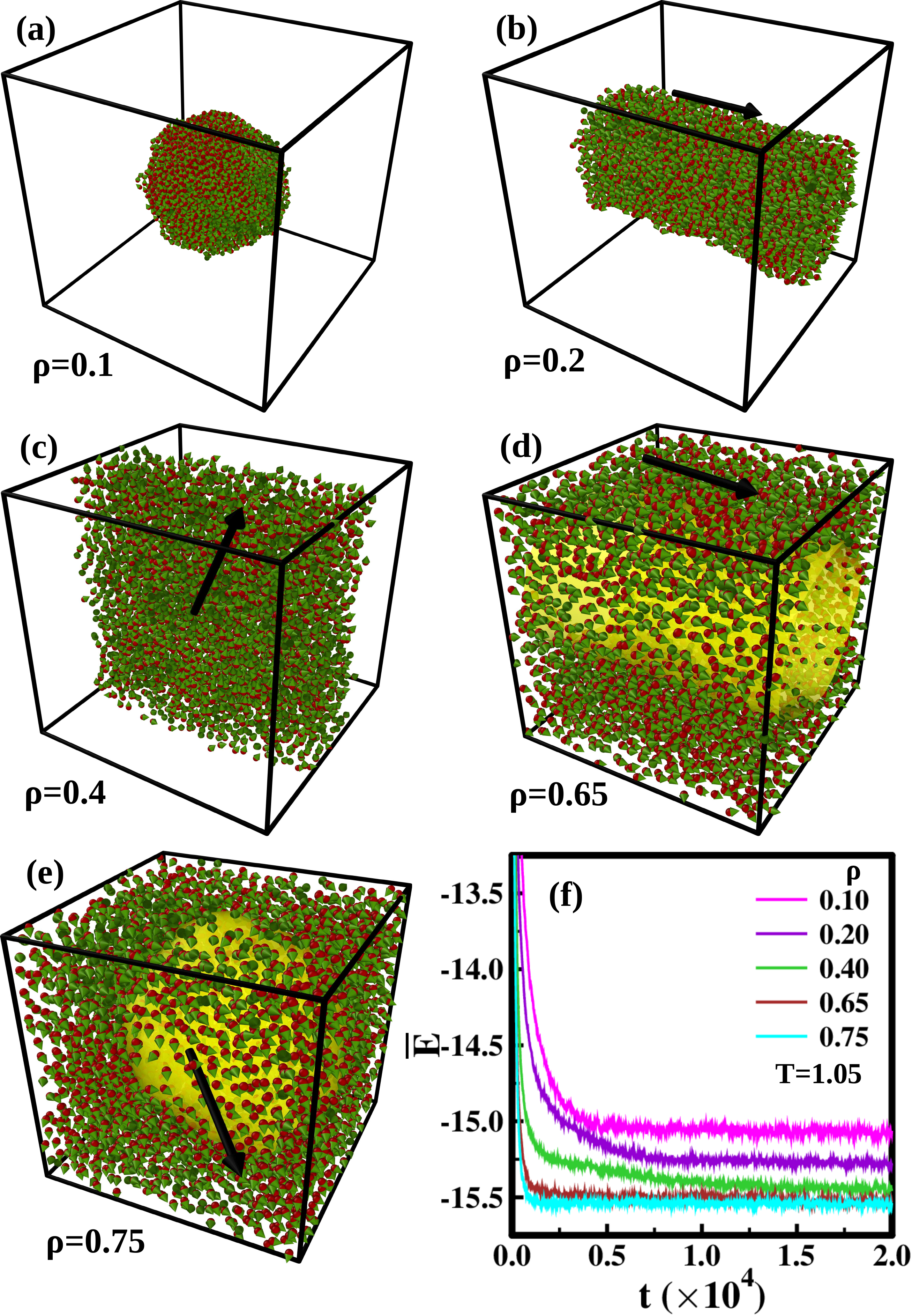}
    \caption{Typical asymptotic morphology shapes that are observed after a quench from $T=5.0$ to $T=1.05$: (a) sphere ($\rho=0.1$),  (b) cylinder ($\rho=0.2$), (c) planar slab ($\rho=0.4$), (d) cylindrical bubble ($\rho=0.65$), (e) spherical bubble ($\rho=0.75$). The yellow color in (d) and (e) indicates the hollow region. The large black arrows represent the direction of the average magnetization ${\bf M}$. (f) Evolution of the ensemble-averaged total energy $\bar{E}$ vs. $t$ for the morphologies in (a)-(e).}
\label{2}
\end{center}
\end{figure}
So what are the long-time structures that are observed in the SM fluid at different values of $\rho$ and $T$? The system is then left to evolve for long times $\sim O(10^6)$ after the quench till when the energy fluctuations are tiny as compared to the average energy. Figure~\ref{2} shows the emergent morphologies at $T=1.05$ for representative values of density: (a) $\rho=0.1$: sphere; (b) $\rho=0.2$: cylinder; (c) $\rho=0.4$: slab; (d) $\rho=0.65$: cylindrical bubble; (e) $\rho=0.75$: spherical bubble. Figure~\ref{2}(f) shows the evolution of the energy $\bar{E}(t)$ vs. $t$, for each value of $\rho$. The bar indicates an average over 20 different initial states. We repeat this exercise for all the points in Fig.~\ref{1}(a), and indicate the emerging structures by different colors: sphere (magenta), cylinder (violet), slab (green), cylindrical bubble (brown), and spherical bubble (cyan). The data points in orange near the binodal line are rare at an early stage of phase separation. Because of the small quantity of the liquid state, the form of the condensate is nearly isotropic. The dashed lines separating the structural phases are a guide to the eye, and roughly indicate the regions where the indicated structures will be observed. (The precise determination of these boundaries will require extensive free energy computations which is beyond the scope of this paper.) 

There are some important points that should be noted in the context of the above morphologies: 
(i) It should be pointed out that the chain formation and ferromagnetic order in the asymptotic morphologies is a consequence of the uninterrupted long-range dipole-dipole interactions realized in our simulations due to the imposition of periodic boundary conditions and the Ewald summation. (With short range interactions on the other hand, an antiferromagnetic alignment of dipolar chains is energetically favorable.) (ii) The asymptotic structures assume shapes with minimum surface energy. This can be checked by evaluating the fraction of the liquid (say, $x=V_l/V$) at any value of $\rho$ and $T$ using Gibb's lever rule. Simple algebra then provides the structures with the least surface area for specific values of $\rho$ and $T$. These evaluations for the morphologies in Fig.~\ref{2} have been provided in Appendix \ref{appen} in Table~\ref{table1}. The size dependence of these structures on $\rho$ and $L$ is also provided in Table~\ref{table3}. (iii) The critical quench and the region close to it, indicated by the green squares in Fig.~\ref{1}(a), yields a slab of dipoles. It is interesting to note that the structures formed for density intervals on either side exhibit complementary assemblies, e.g., cylinder-cylindrical bubble (violet and brown squares) and sphere-spherical bubble (magenta and cyan squares). We mention here that the cylinder (and the complementary cylindrical bubble) can have the long axis along any one of the edges ($x$, $y$, or $z$) of the simulation box. Similarly, the slab width could lie in any of the planes ($xy$, $yz$, or $xz$). (iv) The structures formed in the nucleation region (magenta and cyan) are isotropic while those in the spinodal region (violet, green, and brown) are anisotropic. We will see in Sec.~\ref{sec:4} that the growth laws in the two regimes, which lead to these structures, are also distinct.  

%%%%%%%%%%%%%%%%%%%%%%%%%%%%%%%%%%%%%
\begin{figure}
    \begin{center}
    \includegraphics[width=1\linewidth]{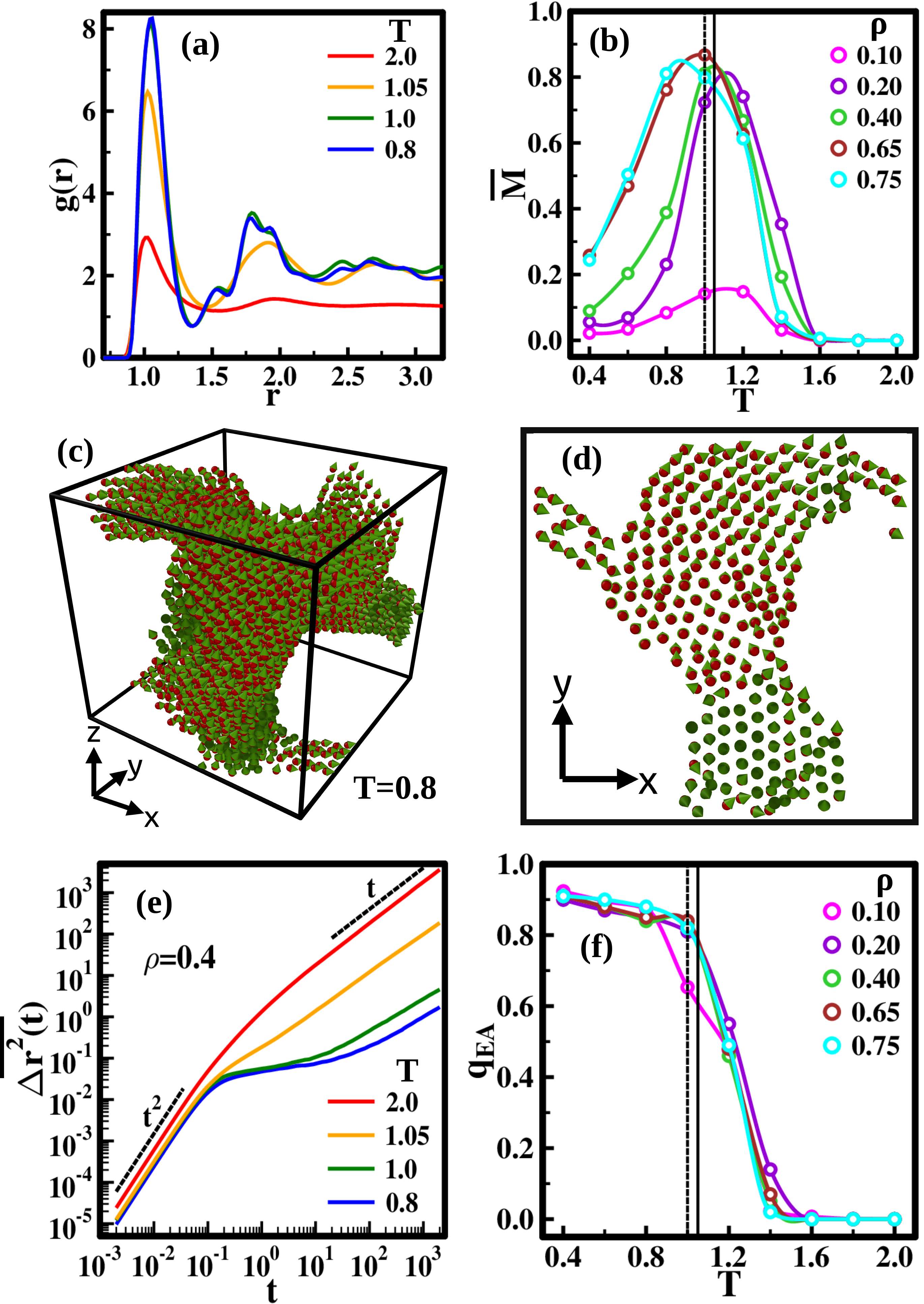}
    \caption{(a) Plot of the PCF for the asymptotic morphologies at $\rho$ = 0.4 at $T$ = 2.0 (isotropic state), $T=1.05$ (liquid state), $T$ = 1.05 (solid-state) and $T_s$ = 1.0 (frozen state). (b) Variation of the ensemble averaged magnetization with temperature, $\bar{M}$ vs. $T$, for $\rho$ = 0.1, 0.2, 0.4, 0.65, and 0.75. (c) A typical frozen morphology ($\rho$ = 0.4) at temperature $T$ = 0.8 and (d) the corresponding $xy$-projection. (e) Mean square displacement for $T$ =2.0, 1.05, 1.0, and 0.8. (f) Variation of the Edwards-Anderson order parameter with temperature, $q_{EA}$ vs. $T$, at specified densities. The dashed and solid lines in (b) and (f) are at T = 1.0 (solid state) and $T=1.05$ (liquid state).}
    \label{3}
\end{center}
\end{figure}

%%%%%%%%%%%%%%%%%%%%%%%%%%%%%%%%%
\begin{figure}
    \begin{center}
		\includegraphics[width=1\linewidth]{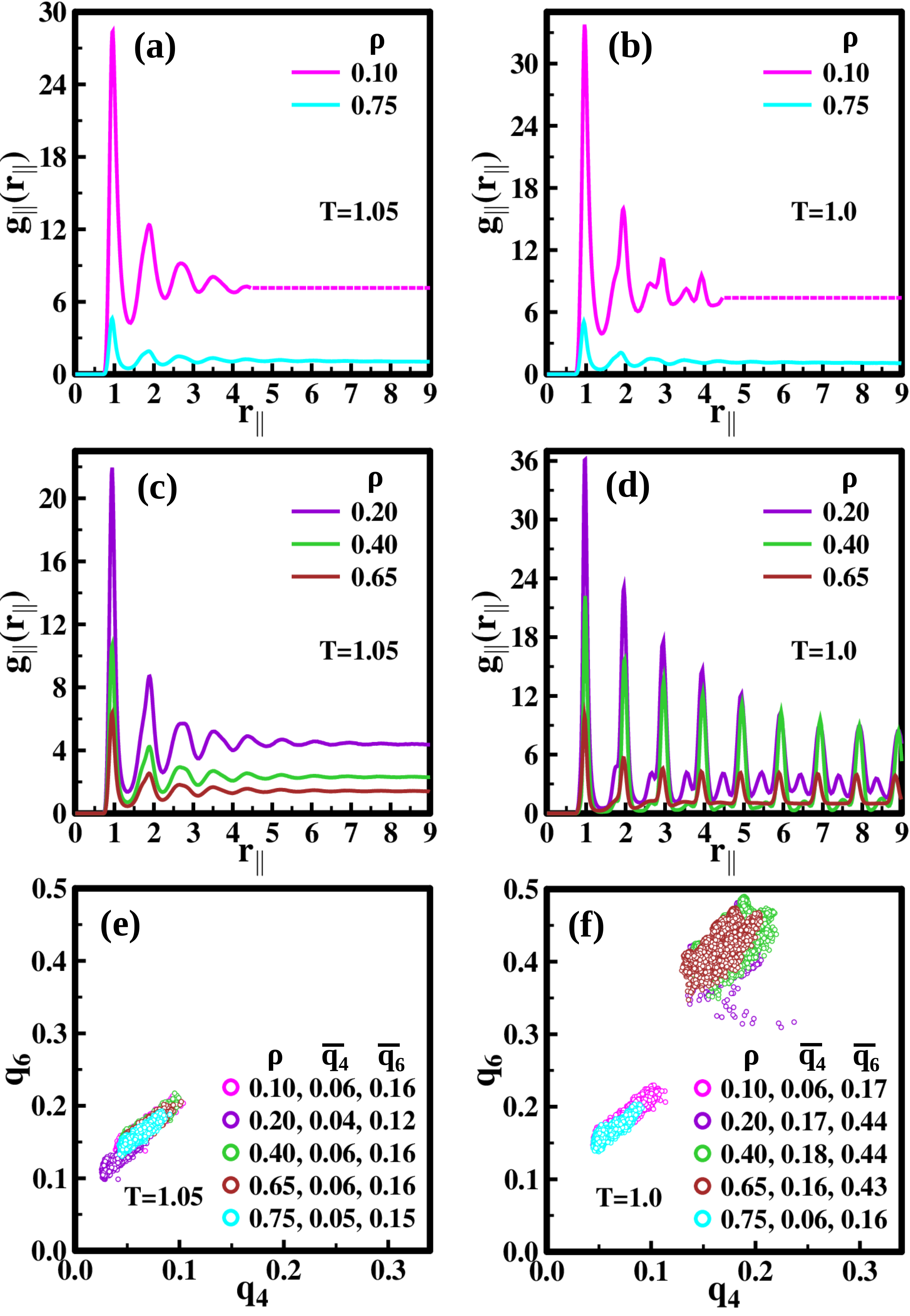}
    \end{center}
    \caption{Plot of longitudinal PCF along ${\bar M}$ in the nucleation region for $\rho=0.1$, and $0.75$ at (a) $T=1.05$ and (b) $T=1.0$. Corresponding PCF in spinodal region for densities $\rho=0.2$, $0.4$, and $0.65$ at (c) $T=1.05$ and (d) $T=1.0$. The scatter plot of local BOP $q_4$ and $q_6$ for $\rho=0.1$, $0.2$, $0.4$, $0.65$, and $0.75$ at (e) $T=1.05$ (liquid state) and (f) $T=1.0$ (solid state). The average values of $q_4$ and $q_6$ for each density are also indicated.}
    \label{4}
\end{figure}

%%%%%%%%%%%%%%%%%%%%%%%%%%%%%%%%%
\begin{figure}
    \centering
        \includegraphics[width=1\linewidth]{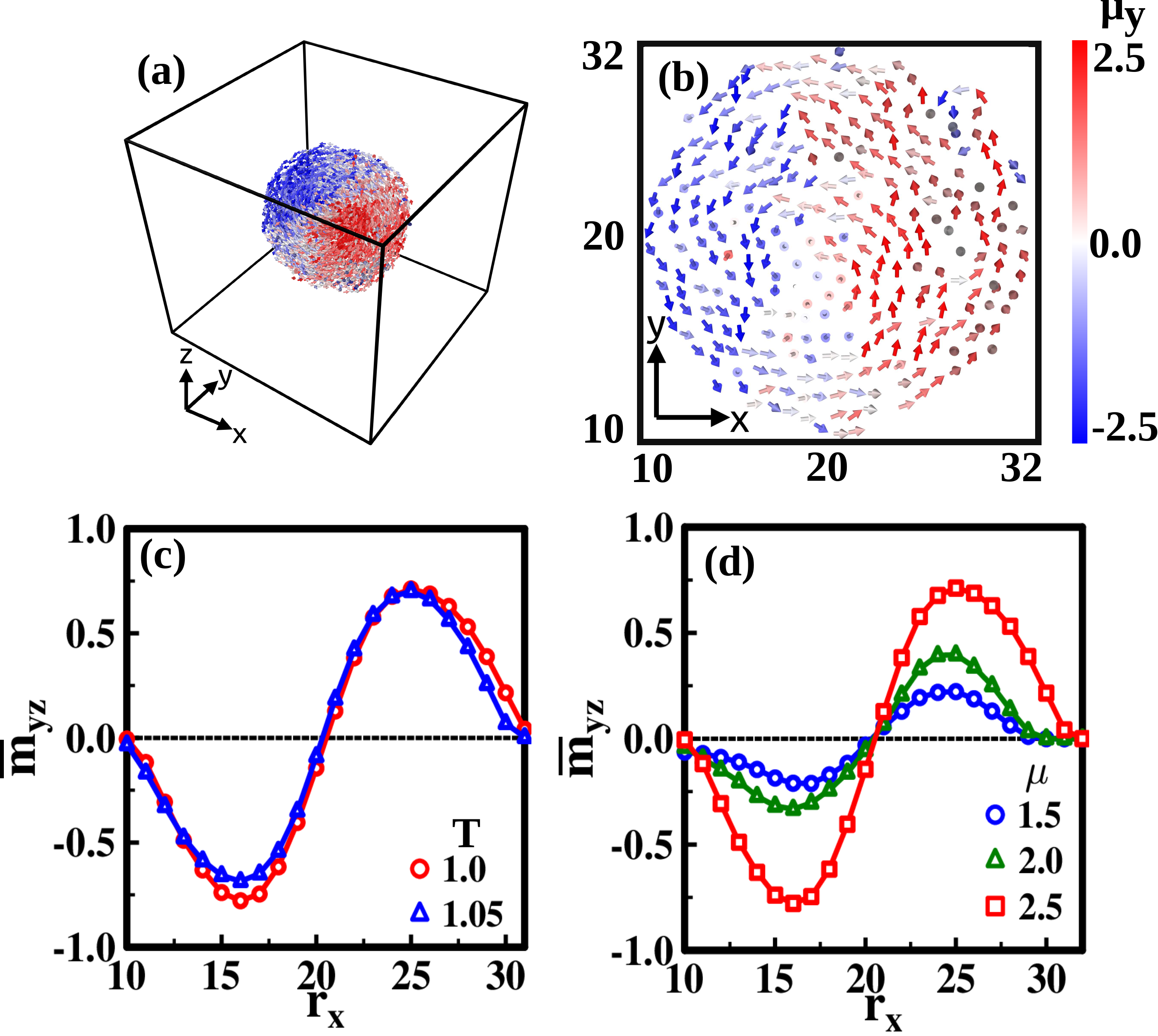}
    \caption{(a) Morphology of the isotropic sphere for $\rho=0.1$ at $T=1.0$ (solid state). The colors indicates the magnitude of the $y$-component of dipole moments $\mu=2.5$, and can be read from the color bar. (b) Corresponding $xy$-projected view of morphology. (c) Variation of average magnetization $\bar{m}_{yz}$ vs. $r_x$ at $T=1.05$ (liquid state)) and $T=1.0$ (solid state). (d) Corresponding comparison for dipole moments $\mu=1.5$, $2.0$, and $2.5$.}
    \label{5}
\end{figure}

Let us investigate the physical state of the condensates as a function of $T$. At each of the points shown in Fig.~\ref{1}(a), we evaluate the PCF $g(r)$ vs. $r$ using Eq.~(\ref{PCF}) to check for the gas, liquid, or solid phase. Figure~\ref{3}(a) shows a prototypical evaluation for $\rho=0.4$ for four values of temperature $T =$ 2.0, 1.05, 1.0 and 0.8. As discussed in Sec.~\ref{sec:PCF}, $g(r)=1$ for the ideal gas, and any deviations imply correlations due to the inter-particle interactions. In Fig.~\ref{3}(a), the evaluation at $T=2.0$ shows the signature of the gas phase. For $T=1.05$, $g(r)$ exhibits a large peak at small-$r$, signifying nearest neighbor correlations followed by small oscillations typical of the liquid phase. The evaluations for $T=1.0$ and 0.8 have the characteristics of the solid phase with several new peaks indicating spatial correlations at values of $r$ that correspond to the {\it lattice spacing}. The development of magnetic order at different temperatures, $M(T)$ vs. $T$,  can be seen in Fig.~\ref{3}(b) for $\rho=$ 0.1, 0.2, 0.4, 0.65 and 0.75. As the temperature is reduced, the dipole-dipole interactions dominate and the magnetization builds up due to the formation of chains of dipoles which co-align parallel to the surface. At very low temperatures, the magnetization goes down presumably due to freezing of the magnetic moments. To understand this aspect, Fig.~\ref{3}(c) shows a typical snapshot at $T= 0.8$ for $\rho = 0.4$ at the latest time $t=10^6$ in our simulation. As anticipated, the system gets stuck in a metastable state that lacks long-range spatial as well as magnetic order. (The expected asymptotic structure for this density is a slab.) This is further emphasized in the $xy$-slice through the center ($z = L/2$) shown in Fig.~\ref{3}(d). To confirm the nature of this phase, we evaluate the mean square displacement $\Delta r^2(t) = \langle \left({\bf r}(t) - {\bf r}(t_0)\right)^2\rangle$ of the dipoles. Figure~\ref{3}(e) shows $\Delta r^2(t)$ vs. $t$ for $T =$ 2.0, 1.05, 1.0 and 0.8. At higher temperatures ($T=2.0$, 1.05), the dipoles exhibit ballistic diffusion, while there is a clear plateau at lower values signifying trapping of dipoles. Further insights can also be obtained from the evaluation of the spin glass Edwards-Anderson order parameter defined by Eq.~(\ref{qEA}). Figure~\ref{3}(f) shows $q_{\text{EA}}$ vs. $T$ for the specified values of $\rho$. The low-$T$ condensates exhibit $q_{\text{EA}} \rightarrow 1$ whereas $M \rightarrow 0$ [Fig.~\ref{3}(b)], which is a characteristic signature of glassy order. Our exercise thus allows us to classify the condensates in the $\rho-T$ plane to be in the liquid above $T_l\simeq 1.05$ [unfilled squares in Fig.~\ref{1}(a)], solid state with quasi-long range order close to $T_s\simeq 1.0$ [filled magenta, purple, green, brown and cyan squares in Fig.~\ref{1}(a)], and a frozen state for $T\leq 0.9$ [filled blue squares in Fig.~\ref{1}(a)]. This phase, with particles in random locations and random orientations, has been observed in experiments \cite{Nakamae2009, Jonsson1995, Dziaugys2010, Maignan2012, Vinod2020} as well as computations \cite{Andresen2014, Ayton1997, Berthier2011, Berthier2023}. The reported slow relaxation and aging have led to the frozen disorder being called spin glass \cite{Andresen2014}, super spin glass \cite{Nakamae2009}, dipolar glass \cite{Ayton1997}, structural glass, and sometimes simply the frozen ferrofluid \cite{Jonsson1995, Vinod2020}. However, even to provide the correct nomenclature, a careful investigation is required to understand the development of this phase and the ferro to glass phase transition if any.

Next, let us understand the circumstances that lead to the large magnetic order in the structures obtained at $T=1.05$ and $1.0$. We evaluate the directional pair PCF $g_{\parallel}(r_{\parallel})$ vs. $r_{\parallel}$ along $\hat{m}$, the direction of anisotropy. The first row of Figs.~\ref{4} show this evaluation for $\rho =$ 0.1 (sphere) and 0.75 (spherical bubble) at (a) $T=1.05$ and (b) $T = 1.0$. The peaks are sharper and more in number for the solid phase. These features are pronounced in the corresponding evaluations shown in Figs.~\ref{4}(c) and 4(d) for the anisotropic structures obtained for  $\rho =$ 0.2 (cylinder), 0.4 (slab), and 0.65 (cylindrical bubble). Next, we evaluate the BOP using Eq.~(\ref{q1}) to obtain information about the local neighborhood of a particle in the condensate. The evaluations of $q_4$ and $q_6$ corresponding to different values of $\rho$ are shown in Figs.~\ref{4}(e) and 4(f) for $T=1.05$ and 1.0. The average values of $q_4$ and $q_6$ for each $\rho$ are also indicated. The evaluations for $T=1.05$ clearly suggest that there is no local spatial order in the liquid state although there is magnetic order. In the solid state for $T=1.0$, it is interesting to note that while the isotropic structures do not exhibit crystalline order, the anisotropic structures are close-packed with FCC order dominating over HCP.

%%%%%%%%%%%%%%%%%%%%%%%%%%%%%%%%%
\begin{figure}
    \begin{center}
		\includegraphics[width=1\linewidth]{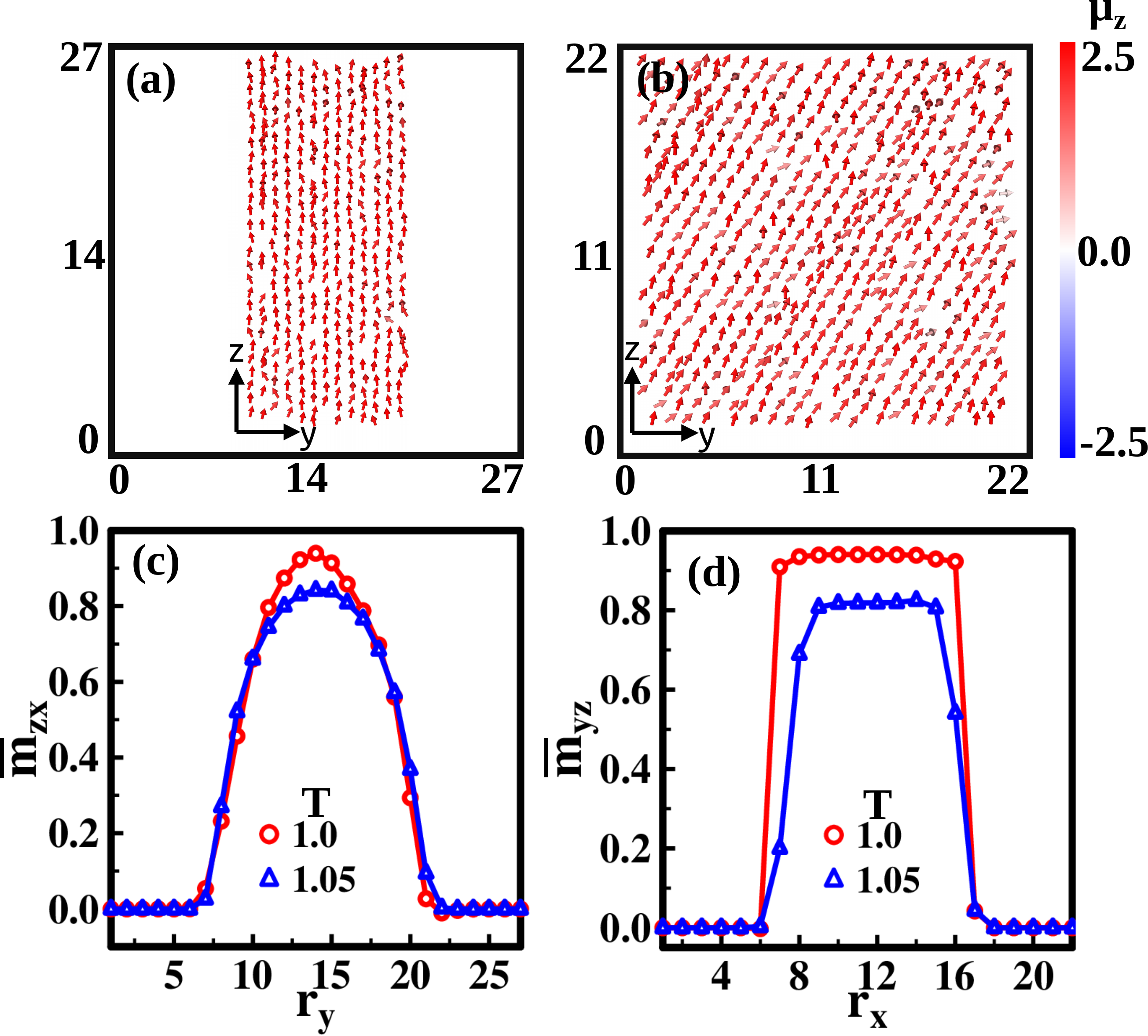}
    \end{center}
    \caption{Slices of (a) cylinder ($\rho=0.2$) and (b) slab ($\rho=0.4$) at temperature $T=1.0$. The color coding indicates the magnitude of $\mu_z$. (c) Variation of the average magnetization $\bar{m}_{zx}$ vs. $r_y$ for cylinder at $T$ = 1.05 and 1.0. (d) Variation of $\bar{m}_{yz}$ vs. $r_x$ for the planar slab at $T=1.05$ and 1.0.} 
    \label{6}
\end{figure}

%%%%%%%%%%%%%%%%%%%%%%%%%%%%%%%%%
\begin{figure}
    \begin{center}
		\includegraphics[width=1\linewidth]{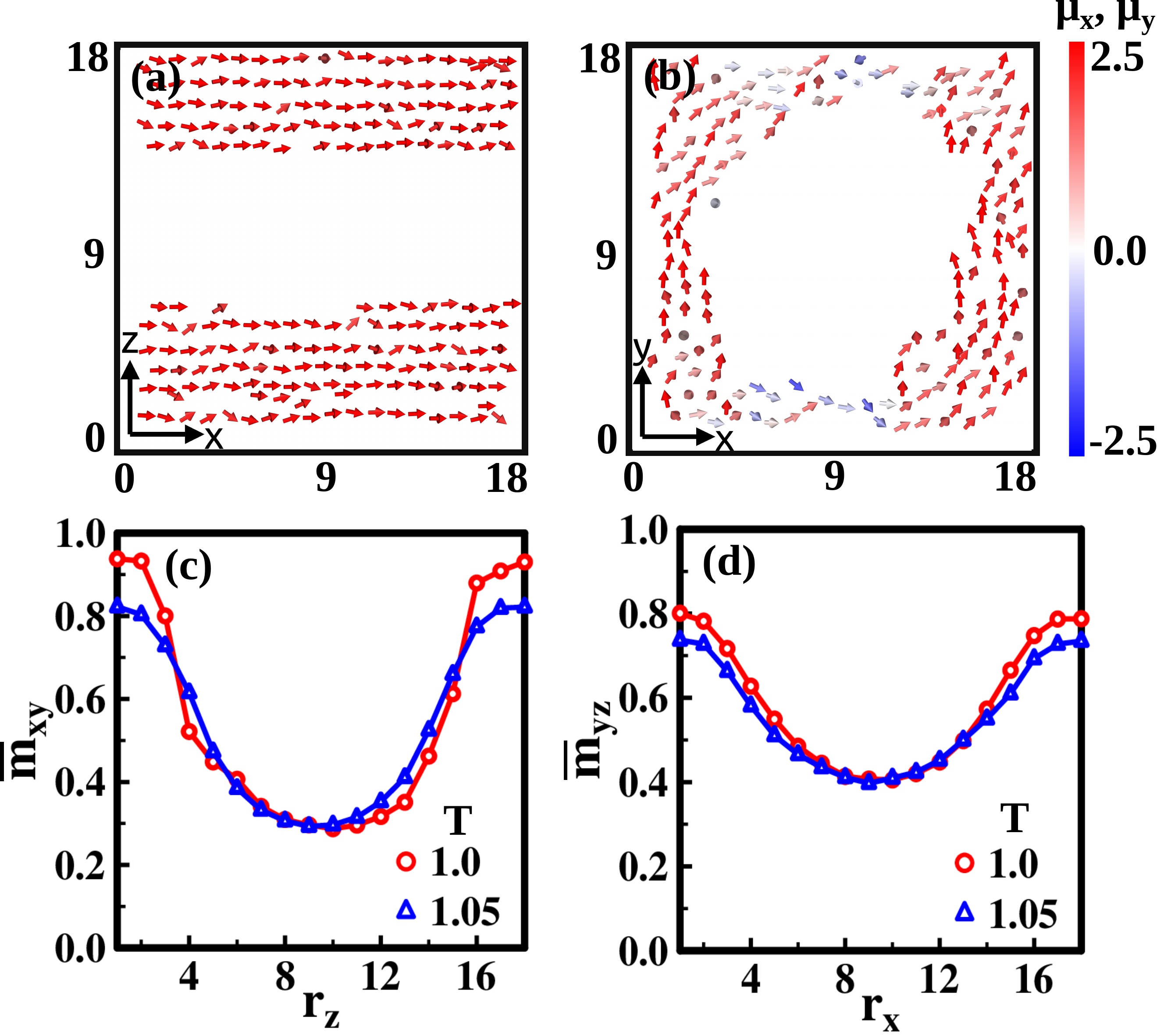}
    \end{center}
    \caption{Slices of (a) cylindrical bubble ($\rho=0.65$) and (b) spherical bubble ($\rho=0.75$) at $T=1.0$. The color coding indicates the magnitude of  $\mu_x$ in (a) and $\mu_y$ in (b). (c) Variation of $\bar{m}_{xy}$ vs. $r_z$ for the cylindrical bubble for $T=1.05$ and 1.0. (d) Variation of $\bar{m}_{yz}$ vs. $r_x$ for the spherical bubble for $T=1.05$ and 1.0.}
    \label{7}
\end{figure}

We now concentrate on unearthing some interesting magnetic properties that develop in the condensates due to the interplay of the surface energy and the dipole-dipole interactions. When the latter dominate over the disordering effects of temperature, the magnetic moments form long chains that align along the surface, and this leads to some unusual characteristics. Figure~\ref{5}(a) shows the components of the magnetic moment along the magnetization axis $\hat{m}$ (or, say, $y$-axis) for the isotropic sphere. The magnitude of dipole moments $\mu_y$ is provided by the adjacent color bar. Figure~\ref{5}(b) shows the corresponding slice for clarity in the alignment of magnetic moments. They co-align along the surface. Figure~\ref{5}(c) shows the variation of the average magnetization in the $yz$ slice ($\bar {m}_{yz}$) as we move along the $x$ direction for both $T=1.05$ (liquid) and $T=1.0$ (solid). The organization of the magnetic moments in the sphere is unusual:  As is evident from the red and blue sections in  Fig.~\ref{5}(a), we have a  {\it magnetic Janus sphere} \cite{Ren2014} composed of two hemispherical domains with opposite magnetic orientations. Further, there is no significant change in the $\bar {m}_{yz}$ vs. $r_x$ behavior for the condensate in the solid or liquid state. Figure~\ref{5}(d), which shows the magnetization scans for different values of $\mu$, suggests that the magnetization is only enhanced for larger values of the magnetic moment.

Figures~\ref{6}(a) and 6(b) show the cross-sections corresponding to the anisotropic cylinder and slab for $T=1.0$. The moments align along the surface to form long chains. The average magnetization of the slices along specified directions is shown in Figs.~\ref{6}(c) and 6(d) for both $T=$ 1.0 and 1.05. The cylinder and slab morphologies thus provide uniformly magnetized self-assemblies. Their size can be manipulated by the density $\rho$ and the size $L$ of the simulation box; please see Appendix \ref{appen} for the details providing the dependence on parameters. As an illustration, for $\rho = 0.2$, $L = 20$, the radius of the cylinder $r_c = 5.41$. On the other hand, $L = 40$ results in $r_c = 10.83$, while $L =80 $ results in $r_c = 21.65$. We show the slices of the cylindrical and spherical bubbles in Figs.~\ref{7}(a) and 7(b). The average magnetization within the bubbles is also shown in Figs.~\ref{7}(c) and (d). As with the other condensates, the size of the bubbles can also be tailored by adjusting $L$ and $\rho$, please see Appendix \ref{appen} for useful information. 
%%%%%%%%%%%%%%%%%%%%%%%%%%%%%%%%%
\begin{figure*}
    \centering
	\includegraphics[width=1\textwidth]{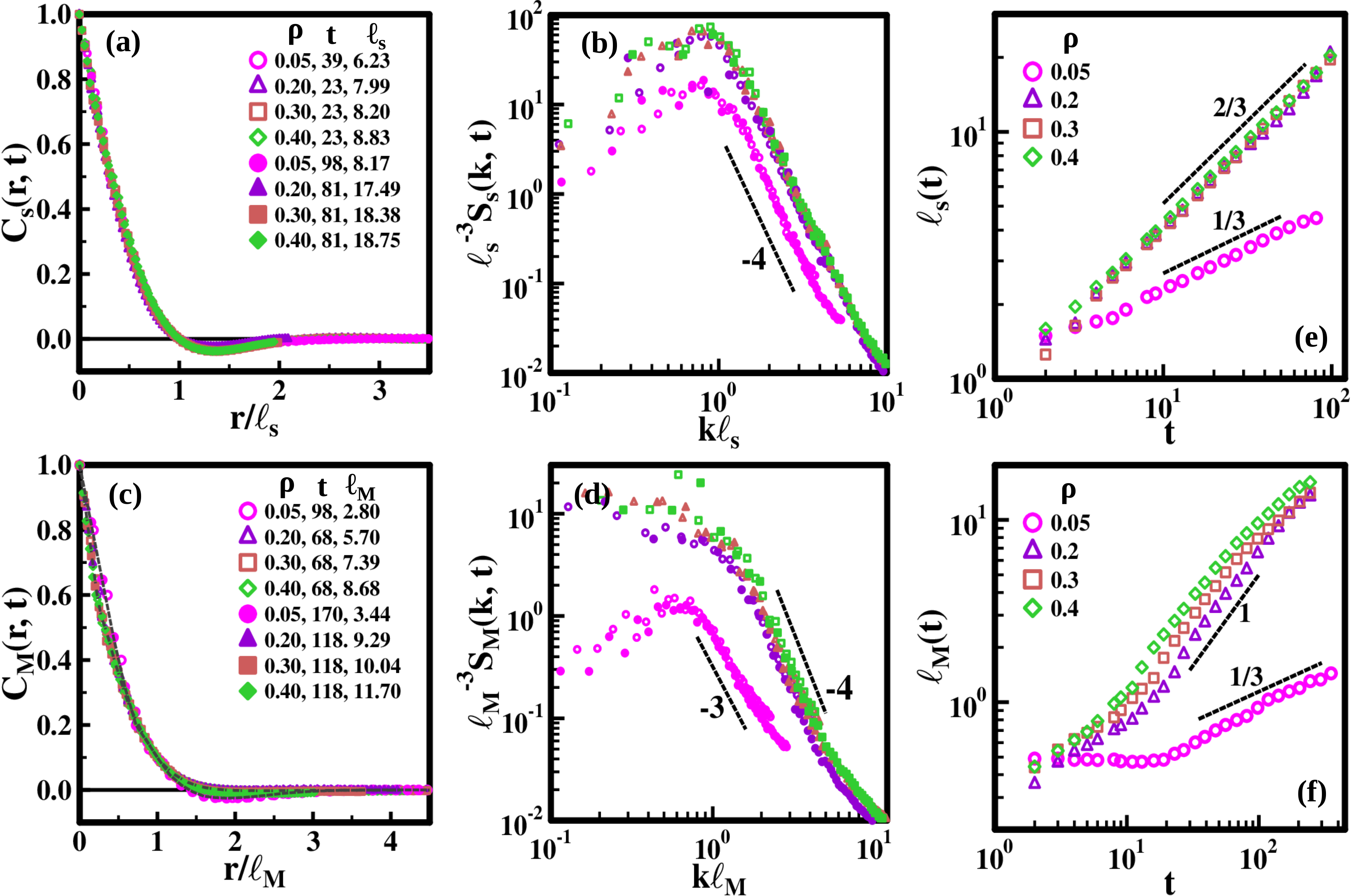}
    \caption{(a) Plots of scaled spatial correlation function $C_s(r, t)$ vs. $r/\ell_s$ for specified values of $\rho$ and $t$. (b) The corresponding scaled structure factor $S_s(k,t)$ vs. $k\ell_s$ on a log-log scale. The dashed line denotes the relevant Porod tail.  (c) Plots of scaled magnetic correlation function $C_M(r, t)$ vs. $r/\ell_M$ for densities and time as mentioned alongside the plot. The dashed line indicates the scaling function for $\rho=0.05$, while the dash-dotted line is for the other values of $\rho$. (d) The corresponding scaled structure factor $S_M(k,t)$ vs. $k\ell_M$ on a log-log scale. The dashed line denotes the relevant Porod tail. (e) The characteristic length scale $\ell_s(t)$ vs. $t$ on the log-log scale for nucleation ($\rho=0.05$) and spinodal ($\rho=0.2$, $0.3$, $0.4$) regimes. (f) The characteristic magnetic length scale $\ell_M(t)$ vs. $t$ on the log-log scale for the nucleation and spinodal regimes. The dashed lines with specified slopes are a guide to the eye. The length scale data has been shifted for clarity. }
    \label{8}
\end{figure*}

\section{Non-equilibrium evolution}
\label{sec:4}
Finally, we also study the non-equilibrium evolution to understand the dominant transport at different densities. As discussed in Sec.~\ref{sec:cor}, a useful tool in this context is the equal-time correlation function defined by Eq.~(\ref{corrfn}). The evolving morphologies develop spatial as well as magnetic order, so it is essential to evaluate the spatial correlation lengthscale $\ell_s$ as well as the magnetic correlation lengthscale $\ell_M$. For this evaluation, we map the continuum system onto a spin-lattice by discretizing the volume $V$ into sub-boxes of size $2^3$. (Our results do not depend on the size of the sub-box.) A sub-box $i$ centered at $\vec{r}_i$ with density $\rho_i>\rho$ is identified as a liquid phase with $\psi_s(\vec{r_i})=1$. On the other hand, $\rho_i<\rho$ is identified as the gas phase with $\psi_s(\vec{r}_i)=-1$. For the magnetic order in the liquid phase, the order parameter $\psi_{M}(r_i)$ is the average dipole moment of the particles in sub-box $i$. Figure~\ref{8}(a) shows the scaled correlation function $C_s(r, t)$ vs. $r/\ell_s$ for $\rho$ = 0.05, 0.2, 0.3, 0.4 and specified values  $t$. (We do not have data for higher densities because the required computational cost is beyond our available resources.) The average domain length for the liquid phase $\ell_s$ is defined as the first zero crossing of the correlation function $C_s(r, t)$. This value for each data set has been specified in Fig.~\ref{8}(a). The small dip in $C(r)$ is characteristic of periodic modulations in bi-continuous morphologies \cite{Bray2002, Puri2009}.  The system exhibits dynamical scaling for all values of $\rho$ indicating the presence of a unique lengthscale. The data also scale for the different values of $\rho$ which span the nucleation as well as the spinodal regime. The corresponding scaled structure factor $\ell_s^{-3}S_s(k,t)$ vs. $k\ell_s$ shown in Fig.~\ref{8}(b) has a Porod tail, $S_s(k)\sim k^{-4}$ due to scattering from smooth GL interfaces. 

Similarly, Fig.~\ref{8}(c) shows scaled magnetic correlations  $C_M(r, t)$ vs. $r/\ell_M$. Note that the dip in the correlations is observed for $\rho =0.05$ in the nucleation regime but not for $\rho = $ 0.2, 0.3, 0.4 in the spinodal regime. The average magnetic domain size $\ell_M$ is also provided for each value of $\rho$ and $t$. It is defined as 0.1 of the maximum value of correlation function $C_M(r, t)$. This data also exhibits dynamical scaling, but the scaling functions are distinct for the nucleation regime and the spinodal regime. The dash-dotted line shows the scaling function for $\rho = $ 0.2, 0.3, 0.4. The dashed line guides the data collapse corresponding to $\rho=0.05$. Further, this data also shows a dip seen in systems described by a conserved order parameter and is consistent with our observation of the magnetic Janus sphere. The corresponding scaled structure factor $\ell_M^{-3}S_M(k,t)$ vs. $k\ell_M$ is shown in Fig.~\ref{8}(d). Interestingly, $S(k)\sim k^{-3}$ for $\rho=0.05$ is indicative of the scattering off the 2-$d$ interface separating the hemispherical domains of up spins and down spins. For higher densities in the spinodal regime, the data exhibits a Porod tail $S(k)\sim k^{-4}$. It should be mentioned that for an $n$-component order parameter, the tail is expected to obey the generalized Porod law: $S(k)\sim k^{-d+n} \equiv k^{-6}$ characteristic of scattering from monopoles and hedgehogs \cite{Bray2002, Puri2009}. In our simulations, the morphologies have smooth GL interfaces. Consequently, the interfacial scattering $S_M(k)\sim k^{-4}$ dominates.  

Let us further quantify the growth of spatial and magnetic correlations in the condensates. Figure~\ref{8}(e) shows $\ell_s(t)$ vs. $t$ for $T=1.05$ at densities $\rho=$ 0.05, 0.2, 0.3, and 0.4. The dashed lines with slopes 1/3 and 2/3 are guides to the eye. The exponent $1/3$ captures the growth in the nucleation regime indicative of diffusive growth. The bi-continuous morphologies in the spinodal regime follow a $2/3$ growth law. Thus the fluid inertia overpowers the capillary and viscous forces right from the onset. This hydrodynamics-driven inertial growth has been elusive in MD simulations, and was observed in our recent study which focused on the phase ordering in the spinodal regime \cite{Singh2023}. We refrain from reproducing the details to avoid repetition. Figure~\ref{8}(f) shows the corresponding variation of $\ell_M$ vs. $t$ for the growing magnetic condensates. The growth of magnetic correlations is delayed as compared to the spatial correlations, indicating that they are triggered by condensation. The dashed lines guiding the eye suggest that the growth exponent is $1/3$ in the nucleation regime. (Larger system sizes will be required to observe a cleaner growth law.) Here, the magnetic order parameter exhibits conservation due to the presence of two oppositely magnetized hemispheres. The growth exponent $\sim 1$ for the bi-continuous morphologies in the spinodal regime is consistent with observations in dipolar solids, as discussed in our earlier study \cite{Singh2023}.

%%%%%%%%%%%%%%%%%%%%%%%%%%%
\section{Conclusion}
\label{sec:5}
Let us conclude with a summary and discussion of our results. We have performed extensive MD simulations to understand the asymptotic phases and non-equilibrium behavior of the SM which consists of LJ particles carrying an embedded point dipole. This simplest polar counterpart of the LJ fluid exhibits GL coexistence, and is a representative model to study ferrofluids, magneto-rheological fluids, electro-rheological fluids, dipolar fluids, etc. These systems have promising applications, as they exhibit the dual properties of fluidity and magnetism. We performed quenches from a high temperature ($T>T_c$) homogeneous gas phase into the coexistence region ($\rho-T$ plane, $T<T_c$) and studied the non-equilibrium evolution for long times to obtain the asymptotic morphologies. All simulations were performed using LAMMPS in the NVT ensemble.% using the Langevin thermostat for the {\color{blue}asymptotic} studies and the Nos\'e-Hoover thermostat to capture the hydrodynamics during the non-equilibrium evolution. %The final states are independent of the choice of the thermostat.   

A systematic variation in the $\rho-T$ plane reveals density-dependent features. For quenches in the nucleation regime, the growth of the condensed phase is via diffusive motion of the SM gas particles. The corresponding growth law is the Lifshitz-Slyozov law $\ell(t)\sim t^{1/3}$ characteristic of binary systems described by a conserved order parameter. The typical self-assemblies are isotropic: lower densities yield a compact sphere while higher densities yield the complementary spherical bubble structure. When quenches are in the spinodal region, the formation of the condensates is driven by the overpowering fluid hydrodynamics as revealed by the inertial growth law $\ell(t)\sim t^{2/3}$ right from the onset of phase separation. The self-assemblies are anisotropic with shapes ranging from cylinder, rectangular slab, and cylindrical bubble. The slightly delayed magnetic order in the condensates shows the characteristics of a non-conserved order parameter, with the development of smooth interfaces separating the magnetically ordered domains (condensates) from the coexisting magnetic vapor. The SM fluid thereby exhibits unusual aspects with combine the physics of conserved and non-conserved order parameters in the spatial as well as magnetic order.

The co-alignment of the magnetic moments along the surface imparts magnetic properties to the self-assemblies. For example, the oppositely magnetized hemispheres of the Janus sphere allow for remote manipulation of headed movement and orientation \cite{Campuzano2019}. It can have many applications, ranging from elementary building blocks for larger self-assemblies, active matter, and drug delivery to name a few \cite{Ren2014, Lattuada2011}. Magnetic bubbles have emerged as another class of materials. The confined hollow geometry and pronouncedly curved surfaces induce unique physical properties different from those of flat thin films and solid counterparts \cite{Ye2010, Retsch2011}. The surface modification opens up possible applications in the areas of catalysis, drug-delivery systems, and magnetic photonic crystals \cite{Armenia2022}. Further, the use of magnetic bubbles as memory devices has been established because of their non-volatility and high reliability originating from their robust structure \cite{Stadler1994}. Simple energy calculations using the system density $\rho$ allows us to estimate the dimensions of these structures accurately. Such precision can allow for control of the spatial and magnetic properties that are required for the above applications. %The methodologies and novel observations presented in this work can provide provoke focussed experimental investigations for exciting physics and applications. 

There can be many extensions of our paper with clues given by earlier works in literature. For example, studies ($d=2$) to see pattern formation in mixtures of magnetic and non-magnetic particles, or self-assembly ($d=3$) from complex building blocks such as chains, rings, X and Y shapes, etc. \cite{Kogel2018, Novak2019, Novak2020, Zverev2021, Novikau2023, Islam2003}. It may be interesting to explore the role of composition in binary mixtures ($d=3$) to obtain wrapped spheres, cylinders, and slabs or their Janus counterparts. for varied applications or to identify the co-existing phases and internal organization in self-assembled structures from multi-particle building blocks. Such studies may have implications in proposing functional materials and comprehending cellular organization in bio-inspired self-assemblies for instance. Further, insights into the frozen phase can also unfold mysteries. 

Experiments with ferrofluids have usually been performed using dilute samples ($\rho\lesssim 0.1$) in confined environments which truncate the long-range dipole-dipole interactions \cite{Klapp2005, Ivanov2020, Butter2003, Borin2020, Ye2016, Martinez2021}. At low densities ($\rho \lesssim 0.1 $), long chains of dipoles have been reported in thin samples \cite{Klapp2005, Van2005, Kantorovich2015}.  For higher densities ($\rho = 0.3 $), there are reports of ferromagnetic fluctuations in zero field as seen from static Susceptibility measurements \cite{Mamiya2000} as well as ac susceptibility measurements \cite{Lebedev2019}. These works, however, did not study the shapes of aggregates and magnetic organization therein. Our simulation results, on the other hand, mimic large systems which allow us to see the consequence of long-range interactions. We hope that our paper will initiate experimental strategies that can verify our observations because of their exciting physics and practical utility.

\noindent {\bf Acknowledgements}: AKS acknowledges IIT Delhi for a research fellowship. VB acknowledges SERB (India) for a CORE research project and a MATRICS grant. The HPC facility at IIT Delhi is gratefully acknowledged for computational resources.

\appendix

\section{Surface energy calculations of the asymptotic morphologies}
\label{appen}
We have used the GL coexistence phase diagram of the SM fluid at dipole moments $\mu=2.5$ that is available in the literature \cite{Stevens1995}. %We have extracted the liquid and gas densities ($\rho_g$ and $\rho_l$) as the function of temperature from \cite{Stevens1995}. 
Gibb's lever rule provides the volume fraction of the liquid ($x=V_l/V_0$) in the coexistence region by the following expression:
\begin{equation}
   x(T) = \frac{\rho - \rho_g(T)}{\rho_l(T) - \rho_g(T)}, 
\end{equation}
where $\rho$ is the system density, $V_l$ is the volume of the liquid state while $V_0$ is the volume of the system. We have extracted the gas and liquid densities $\rho_l(T)$ and $\rho_g(T)$ using the GL coexistence phase diagram of the SM fluid for $\mu=2.5$ from ref. \cite{Stevens1995}.

As an example, let us obtain the radius $r$ of the sphere in terms of $x$ using Gibb's lever rule.
\begin{equation}
\label{glr}
   x(T) = \frac{V_l}{V_0} = \frac{\left(4/3\pi r^3\right) }{L^3}. 
\end{equation}
Therefore,
\begin{equation}
\label{radius}
   r = L\Big{(}\frac{3x}{4\pi}\Big{)}^{\frac{1}{3}}, 
\end{equation}
where $L=(N/\rho)^{1/3}$ is the length of the simulation box. The surface area of the sphere $A_s$ = $4\pi r^2$ is then evaluated by substituting for $r$ from Eq.~(\ref{radius}). In Table~\ref{table1}, we provide the evaluations of the radius $r$ (or width $b_{\text{ps}}$ of the planar slab), surface area ($A$), and volume ($V_l$) of the observed asymptotic structures in terms of liquid fraction $x$ and the box dimension $L$ ($V=L^3$). Table~\ref{table2} provides the numerical evaluations of $A$ from our simulations for $T = 1.4$ and representative values of $\rho =$ 0.1, 0.2, 0.3, 0.4, 0.65, and 0.75. The minimum surface energy for a particular value of $\rho$ is colored in red. Table~\ref{table3} shows the effect of increasing $L$  on $r$.

%%%%%%%%%%%%%%%%%%%%%%%%%%%%%%%%%%%%%%%%%
\begin{table}
\caption{Evaluation of radius ($r$), surface area ($A$), and volume ($V_l$) in terms of $x$ and $L$ for sphere (s), cylinder (c), planar slab (ps), cylindrical bubble (cb), and spherical bubble (sb).}
\label{table1}
\begin{ruledtabular}
\begin{tabular}{cccc}
Structures & Radius (r)  & Surface area (A) & Volume ($V_l$) \\
\hline
s & $L(\frac{3x}{4\pi})^{\frac{1}{3}} $  & $4\pi L^2(\frac{3x}{4\pi})^{\frac{2}{3}} $ &  $\frac{4}{3}\pi r^3$\\

c & $L(\frac{x}{\pi})^{\frac{1}{2}}$  &  $2L^2(\pi x)^{\frac{1}{2}} $ & $\pi r^2 L$ \\

ps  &$b=Lx$  &  $2L^2$ &  $L^2b$ \\

cb &  $L\sqrt{\frac{1-x}{\pi}}$  &  $2L^2\sqrt{\pi(1-x)}$ & $L^3-\pi r^2L$ \\

sb & $L\{\frac{3(1-x)}{4\pi}\}^{\frac{1}{3}} $  & $4\pi L^2\{\frac{3(1-x)}{4\pi}\}^{\frac{2}{3}} $ &  $L^3-\frac{4}{3}\pi r^3$\\
\end{tabular}
\end{ruledtabular}
\end{table}

%%%%%%%%%%%%%%%%%%%%%%%%%%%%%%%
\begin{table}
\caption{Surface area calculation for all condensates at $T=1.4$ and specified values of $\rho$. The minimum surface energy for a particular value of $\rho$ is colored in blue.}
\label{table2}
\begin{ruledtabular}
\begin{tabular}{cccccccc}
$\mathbf{\rho}$ & $\mathbf{L}$ & $\mathbf{x}$ & $\mathbf{A_{s}}$ & $\mathbf{A_{c}}$ & $\mathbf{A_{ps}}$ & $\mathbf{A_{cb}}$ & $\mathbf{A_{sb}}$  \\
\hline
0.10 & 34.20 & 0.11 & $\textcolor{blue}{1321.85}$ & 1393.43 & 2339.28 & 3903.99 & 5220.90   \\

0.20 & 27.14 & 0.23 & 1325.69 & $\textcolor{blue}{1244.10}$  & 1473.59  &2295.79 & 3000.62  \\

0.40 & 21.54 & 0.45 & 1323.72 & 1107.12 & $\textcolor{blue}{928.29}$ & 1216.58 & 1501.03 \\

0.65 & 18.33 & 0.74 & 1323.58 & 1020.99 & 671.61 & $\textcolor{blue}{611.48}$ & 668.19   \\

0.75 & 17.47  & 0.85  & 1323.43 & 996.86 & 610.54 & 420.41 & $\textcolor{blue}{418.55}$  \\
\end{tabular}
\end{ruledtabular}
\end{table}

%%%%%%%%%%%%%%%%%%%%%%%%%%%%%%%
\begin{table}
\caption{Effect of increasing the box size $L$ on the radius (width) of the asymptotic morphologies.}
\label{table3}
\begin{ruledtabular}
\begin{tabular}{cccccc}
$\mathbf{L}$ & $\mathbf{r_{s}}$ & $\mathbf{r_{c}}$ & $\mathbf{b_{ps}}$ & $\mathbf{r_{cb}}$ & $\mathbf{r_{sb}}$ \\
\hline
10 & 2.97 & 2.71 & 4.5 & 2.88 & 3.30  \\

20 & 5.95 & 5.41 & 9 & 5.76 & 6.59  \\

40 & 11.89 & 10.83 & 18 & 11.51 & 13.19  \\

80 & 23.78  & 21.65  & 36 & 23.02 & 26.37  \\

160 & 47.57  & 43.30  & 72 & 46.04 & 52.75  \\
\end{tabular}
\end{ruledtabular}
\end{table}

\newpage
%\nocite{*}
\bibliographystyle{apsrev4-1}
\bibliography{biblio}

%%%%%%%%%%%%%%%%%%%%%%%%
\end{document}